\documentclass[twocolumn,showpacs,prl,superscriptaddress]{revtex4}
\usepackage{graphicx}

\begin{document}

\title{Shell filling in closed single-wall carbon nanotube quantum dots}
\author{David H. Cobden}
\email{d.cobden@warwick.ac.uk}
\affiliation{Niels Bohr Institute {\O}rsted Laboratory, Universitetsparken 5, DK-2100 Copenhagen, Denmark}
\affiliation{Department of Physics, University of Warwick, Coventry CV4 7AL, UK}
\author{Jesper Nyg{\aa}rd}
\email{nygard@nbi.dk}
\affiliation{Niels Bohr Institute {\O}rsted Laboratory, Universitetsparken 5, DK-2100 Copenhagen, Denmark}

\date{Dec.\ 6, 2001}

\begin{abstract}
We observe two-fold shell filling in the spectra of closed one-dimensional quantum dots formed in single-wall carbon nanotubes.  Its signatures include a bimodal distribution of addition energies, correlations in the excitation spectra for different electron number, and alternation of the spins of the added electrons.  This provides a contrast with quantum dots in higher dimensions, where such spin pairing is absent.  We also see indications of an additional fourfold periodicity indicative of K-K$^\prime$ subband shells.  Our results suggest that the absence of shell filling in most isolated nanotube dots results from disorder or nonuniformity.
\end{abstract}

\maketitle

Quantum dots are of broad fundamental interest, both because they are ubiquitous in nanoscale systems, and because they enable extensions and generalizations of the physics of atoms, molecules, nuclei, and impurities in solids \cite{ref1}.  Two-dimensional (2D) dots formed in semiconductor heterostructures, with controllable size and contact transparency, have been studied intensively for a decade \cite{ref2}. Recently, thanks to the arrival of new systems including nanoparticles \cite{3}, fullerenes \cite{4} and nanotubes \cite{5}, it has become possible to investigate the influences of geometry, band structure, atomic confirmation, vibrations, and surface chemistry on quantum dots.  As with atoms and nuclei, in quantum dots symmetry-related orbital degeneracies combined with spin and the Pauli principle can lead to electronic shells.  Shells have indeed been observed in small, symmetric 3D and 2D dots, such as warm metal clusters \cite{6} and small vertical semiconductor dots \cite{7}, respectively.  However, shell filling is disrupted if the spatial symmetry is imperfect or if the number of electrons $N$\ is so large that shells overlap in energy.  This explains why the spectra of metal nanoparticles and larger semiconductor dots appear chaotic, lacking (as a result of exchange interactions) even the two-fold shells that could arise from the spin degeneracy of each orbital \cite{8,9}.

One-dimensional (1D) quantum dots are now available \cite{10,11}, in the form of single-wall carbon nanotubes.  In nanotubes the sole orbital symmetry is a two-fold one, corresponding to the K-K$^\prime$ subband degeneracy and resulting from the equivalence of the two atoms in the primitive cell of the graphene structure.  Combined with spin this gives a possibility of four-electron shells.  As yet unpublished reports \cite{12,13} indicate that four-electron periodicity can indeed be seen in `open' nanotube devices, where the contacts are highly transparent and the electron states are not localised on the tube.  Here we report the perhaps more surprising result that even clean `closed' nanotube dots, showing complete Coulomb blockade (CB), can exhibit simple two-electron spin shells in the addition spectrum, as well as the disordered remnants of four-fold shells.  The results imply that, for moderate disorder, exchange corrections are very small in this 1D system, in contrast with the 2D and 3D cases.

\begin{figure}
\includegraphics[width=8cm]{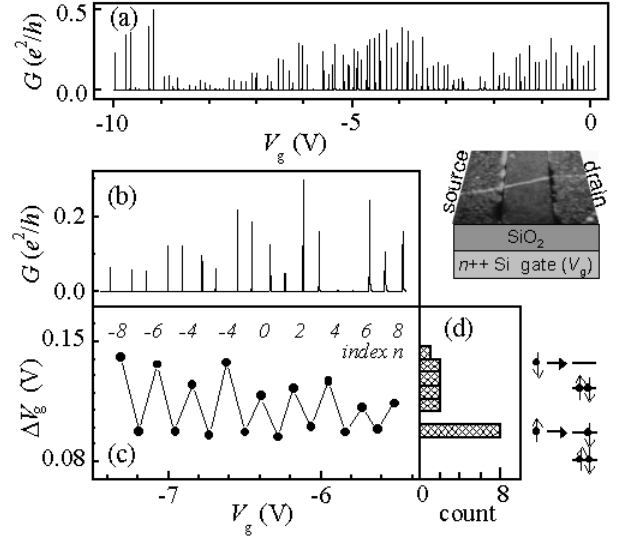}%
\caption{\label{fig1}(a) Conductance vs gate voltage for a nanotube quantum dot at $T = 300$~mK.  The conductance at room temperature was 0.3~$e^2/h$.  (b) Coulomb blockade peaks over a narrower range of $V_g$.  (c) Spacings $\Delta V_g$ of the peaks in (b).  The index $n$\ counts the added electrons relative to an arbitrary zero.  For clarity only even $n$'s are indicated.  (d) Histogram of the data in (c).  The sketches to the right indicate the interpretation of the two peaks in the histogram (see text).  Inset: device structure, incorporating an atomic force microscope image.}
\end{figure}

Our devices were made by evaporating metal (typically 25~nm gold on 5~nm chromium) contacts, patterned by electron beam lithography, on top of nanotubes grown by laser ablation \cite{14} and deposited from a sonicated suspension in dichloroethane onto SiO$_2$.  Their geometry is indicated on the right of Fig.~1.  The separation of the two contacts (source and drain) is $L \sim 300$~nm, and the highly doped silicon below acts as a metallic gate electrode.  In all cases the quantity measured was the dc source-drain current $I$\ with a bias $V$\ applied to the source and the drain grounded.  At low temperature the majority of devices show CB oscillations of the linear conductance $G$ as a function of gate voltage $V_g$, with varying degrees of regularity \cite{15}.  We focus here on one quantum dot, selected from about fifty, which exhibited particularly long and regular sequences of CB peaks, as illustrated in Fig.~1a.

A region containing 18 peaks is expanded in Fig.~1b.  The spacing, $\Delta V_g$, of each pair of adjacent peaks is plotted in Fig.~1c.  We see that it alternates, in such a way that we can index the blockade regions with an integer $n$ which is even for all the larger $\Delta V_g$ values.  In a histogram of $\Delta V_g$ (Fig.~1d) the odd-$n$ values all fall in a bin of width 10~mV centered on~100 mV, while the even-$n$ values are distributed between about 110 and 140~mV.

The quantity $\Delta V_g$ reflects the addition energy, ie, the extra energy required for adding the $N$+1'th electron to the dot relative to the $N$'th \cite{ref1,16}.  Its even-odd alternation naturally suggests two-electron shell filling.  To analyse it further, we start with the constant-interaction (CI) model of a quantum dot.  This model assumes that the total energy depends only on a given set of orbital energies and a single constant interaction parameter $U$, which is the repulsion between any two electrons on the dot.  For odd $N$, the $N$+1'th electron enters the same orbital as the $N$'th, with energy $\epsilon_i$, and the resulting separation of the CB peaks is $\Delta V_g = U/e\alpha$, where $e$ is the electronic charge and $\alpha < 1$ is a capacitance ratio.  For even $N$\ the $N$+1'th electron enters the next orbital, with energy $\epsilon_{i+1}$, and $\Delta V_g= (U + \epsilon_{i+1} - \epsilon_{i})/e\alpha$.  If we identify even $N$\ with even $n$\ in the experiment, then the interpretation of the peaks in the histogram of $\Delta V_g$ is as indicated in the sketches to the right of Fig.~1d.  Using $\alpha = 0.10$ for this device (which can be obtained from the nonlinear measurements discussed below), we then deduce that $U \approx 10$~meV, while $\epsilon_{i+1} - \epsilon_{i}$ is distributed over the range 1--4~meV.  These values are consistent with the usual results for closed nanotube dots \cite{17} that $U/\delta\sim 5$ and $\delta = hv_F/4L$, where $\delta$ is the mean level spacing (assuming the only degeneracy is spin), $v_F \approx 8 \times 10^5$~ms$^{-1}$ is the Fermi velocity, and $\delta \sim 2.7$~meV for $L \sim 300$~nm.

\begin{figure}
\includegraphics[width=8cm]{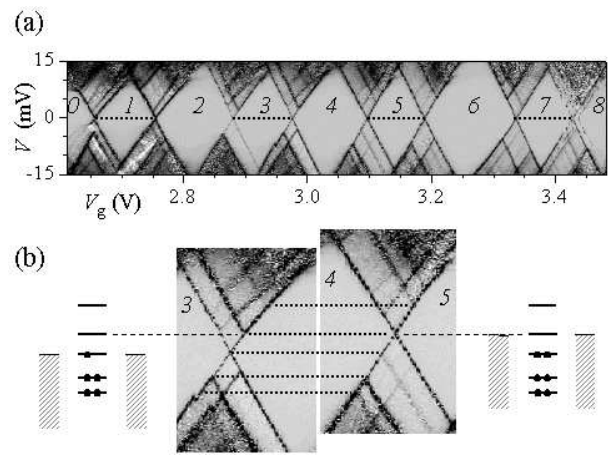}%
\caption{\label{fig2}(a) A greyscale plot (darker = more positive) of ${\rm d}I/{\rm d}V$\ vs $V_g$ and $V$ at $T = 100$~mK and $B = 0$, for the same device on a different cooldown.  The index $n$\ counts electrons added relative to the leftmost diamond. The dotted lines have equal length.  (b) The two central crosses from (a) are shown again, but with a relative vertical displacement.  In the CI model the level spectrum of the dot can be read off from the bias at the points where the excitation lines cross the edge of a diamond.  The resulting spectrum, given by the dotted horizontal lines, is almost the same for both crosses.  The level singled out by the dashed line produces the lowest electron-adding transition on the right and the second lowest on the left, as indicated in the sketches on either side.}
\end{figure}

The CI model predicts that successive CB peaks should show related excitation spectra.  The excitations can be studied \cite{ref2} by making a greyscale plot of ${\rm d}I/{\rm d}V$\ against $V_g$ and the bias $V$, as in Fig.~2a.  In the diamond-shaped blockade regions no current flows and $N$\ is fixed, while within the cross-shaped regions the patterns reflect the spectra of transitions between $N$\ and $N$+1 electrons.  The first prediction is that for a given shell, the allowed transition energies for adding or removing electrons should be identical.  Indeed, we observe that the crosses separated by odd-$n$ diamonds, indicated by dotted lines, bear a clear resemblence (though they are never identical).  The second prediction is that for adjacent shells, the allowed transitions should follow the same pattern but displaced such that the ground state for one corresponds to the lowest excited state for the other \cite{18}.  This prediction also holds true, albeit only approximately, in the nanotube dot.  A particularly close match occurs for the two central crosses in Fig.~2a.  They are shown again in Fig.~2b with the right-hand cross displaced vertically, allowing us to draw horizontal dotted lines which match all the visible electron-adding transitions on both crosses simultaneously.

Furthermore, in the CI model an applied magnetic field $B$\ splits the two spin levels of each orbital by the Zeeman energy.  The result is that most transitions split into pairs, because an electron can be added to an empty orbital, or removed from a full one, with either spin.  The only transitions which do not split are those bordering the odd-$N$\ diamonds, which involve the singly occupied orbital \cite{19}.  We see this pattern whenever the peak pairing can be identified (as well as most the of the time even when it cannot).  For example, Fig.~3 shows the central region from Fig.~2a at $B = 6$~T.  Nearly all the visible transitions are split into pairs, with the exception of those bordering the $n = 3$ and $5$ diamonds.

\begin{figure}
\includegraphics[width=8cm]{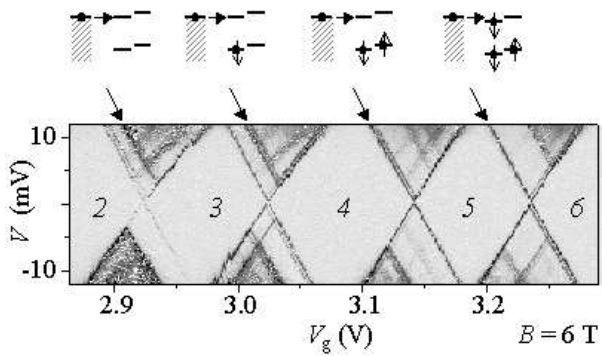}%
\caption{\label{fig3}The central region of Fig.~2a in a magnetic field $B = 6$~T perpendicular to the tube axis and at $T = 100$~mK.  Only the edges of the odd-$n$\ diamonds are not Zeeman split.  This is in agreement with the constant interaction model, as indicated in the sketches above (see text).}
\end{figure}

The essence of the above results is captured in the sketches above Fig.~3, which indicate transitions corresponding to the same orbital on four consecutive CB peaks (crosses).  Such an all-round agreement with the CI model has not been found in semiconductor (2D) or metal (3D) quantum dots.  This is believed to be because of exchange terms, omitted from the CI model, which modify the level spectrum as $N$\ changes \cite{8}.  In 2D and 3D these terms are large enough (relative to $\delta$) to shuffle the order in which states are occupied and preclude spin shell filling.  Our results imply that in 1D dots the opposite is true: exchange effects are small enough for spin shells to exist.

If exchange corrections are indeed small, then the distribution of $\Delta V_g$ for even $n$ in Fig.~1d reflects the distribution of ($N$-independent) level spacings, $\epsilon_{i+1}-\epsilon_{i}$.  For chaotic quantum dots, this distribution is given by random matrix theory (RMT) \cite{20}.  RMT is thought to be appropriate for 2D and 3D dots at large $N$.  However, it is not applicable for a clean 1D dot, which has uniformly spaced levels for all $N$.  In fact, a likely origin of the broad distribution we observe is suggested by scanned-probe measurements \cite{21}, which reveal that nanotubes contain defects.  Each defect backscatters resonantly as a function of gate voltage in a manner dependent on its nature \cite{22}.  A random collection of defects can therefore lead to a complicated spectrum, with a corresponding distribution of level spacings, unique to each nanotube.  It will however probably not randomise the Hamiltonian matrix sufficiently for RMT to be applicable.

In nearly all nanotube dots studied to date (see e.g.\ refs.~\cite{10} and \cite{11}), most of the Coulomb diamonds are distorted, with accompanying complex and variable excitation patterns.  Similar distortions occur in our selected device at some gate voltages, as can be seen in Figs.~4a and b.  The lines of shallower slope in this grayscale plot correspond to a different source-gate capacitance ratio, implying that the geometry of the nanotube dot is not perfectly well defined and can effectively differ between eigenstates.  This too can be explained by imperfections: a resonant defect may leave most electron states delocalized throughout the tube, but near resonance it can effectively divide the tube electronically into smaller dots with different capacitances and very asymmetric couplings to the source and drain \cite{21}. It may be that our device acts as a single quantum dot over large regions of $V_g$ because it contains relatively few defects.  Other types of imperfection are also present, including a nonuniform (and possibly dynamically varying) electrostatic potential along the tube, resulting from the electrode potentials and from nearby trapped charges, which it has been suggested \cite{23,24} may account for the occasional observation of sequences of electrons appearing to enter the tube with the same spin \cite{25}.

\begin{figure}
\includegraphics[width=8.6cm]{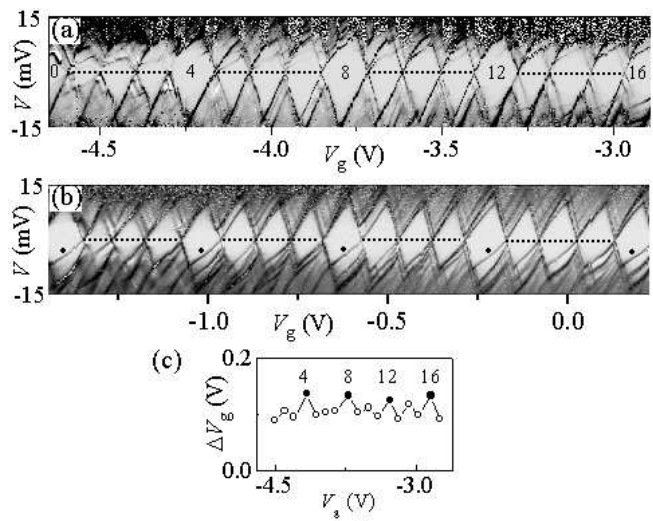}%
\caption{\label{fig4} (a) and (b) Greyscale plots of ${\rm d}I/{\rm d}V$\ vs $V_g$ and $V$ at $T = 300$~mK and $B = 0$, for the same device on a different cool-down. The dotted lines indicate apparent four-fold grouping of the peaks. (c) Peak spacings for (a), with values for $n = 4p$\ plotted as solid circles, where $p$ is an integer and the zero of $n$\ is chosen conveniently.}
\end{figure}

The observation of spin shells prompts the question of whether the K-K$^\prime$ band degeneracy also affects the shell structure.  Combined with spin this might produce a fourfold, instead of a twofold, grouping of the peaks.  No fourfold grouping is obvious in the data of Figs.~1--3.  This does not seem surprising, because the K and K$^\prime$ states are very likely to be mixed, either by defects or at the contacts, lifting their degeneracy.  Indeed, the measured average level spacing only agrees with theory if this degeneracy is assumed lifted (leading to $\delta= hv_F/4L$).  However, consider instead the selections of data in Fig.~4.  In Fig.~4a the peak spacings, plotted in Fig.~4c, are larger for every fourth diamond.  Meanwhile, in Fig.~4b, we see that certain features repeat every four peaks; for example, the pronounced skewing of the ground-state transition line, which we have marked with solid circles in Fig.~4b.  The fact that a fourfold periodicity survives in this complex data could be explained as follows.  Assume that a given ($i$'th) quartet of peaks is associated with wavefunctions having $i$\ nodes along the tube.  The wavefunctions and energies may be disrupted by disorder and exchange.  However, the $i$+1'th quartet will be disrupted in a similar way to the $i$'th, because the addition of one extra node has little consequence as long as the energy is not close to a defect resonance (where the scattering phase varies rapidly with energy).  The result will be a complex pattern which repeats, approximately, every time four electrons are added.

We have noticed a clearer four-fold peak grouping in devices which exhibit strong Kondo resonances of the conductance \cite{26}.  Moreover, very recent studies of similar devices with high contact transparency have shown unmistakable four-fold patterns \cite{12,13}.  With such ``open" contacts the current is never strongly blockaded, and the electronic states are only weakly confined in the tube.  This makes them less sensitive both to disorder and to interactions in the tube, the latter being confirmed by the greatly reduced values of $U/\delta$\ observed.  With nearly ideal contacts, it appears in fact that the charging energy $U$\ can be neglected completely and the device treated as a noninteracting electron waveguide with weak scattering at the contacts only \cite{27}.

The electron system in a nanotube is nevertheless expected to show strong correlations described by Luttinger liquid theory, for which there is a growing body of evidence \cite{17}.  Many intriguing questions remain as to how the correlations relate to the excitation spectrum, the disorder, and the apparently small exchange terms that allow spin shells to be seen.  Nanotube devices containing fewer or no defects, and with improved control of the electrostatic potential, will be required to further our understanding of these finite 1D quantum systems.

In summary, we find that {\em closed}\ one-dimensional quantum dots, unlike their higher dimensional counterparts, can exhibit twofold spin shell filling.  We thank Nick d'Ambrumenil, Karsten Flensberg, Poul Erik Lindelof, Boris Muzykantskii, and Neil Wilson for helpful discussions. The experiments were supported by the Danish Research Councils (SNF, STVF).

\end{document}